\documentclass[aps,prl,twocolumn,showpacs,longbibliography]{revtex4-1}
\pdfoutput=1
\usepackage{amsmath}
\usepackage{amsfonts}
\usepackage{amssymb}
\usepackage{graphicx}
\usepackage{hyperref}
\usepackage{mathtools}
\usepackage{verbatim}
\usepackage{epstopdf}
\usepackage{subfigure}
\usepackage{latexsym}
\usepackage{dcolumn}
\usepackage{epsf}
\usepackage{float}
\usepackage[table]{xcolor}
\usepackage{multirow}
\usepackage{makecell}

\usepackage{color}
\begin{document}

\newcommand{\I} 
	{
		\mathcal I
	}
\newcommand{\A} 
	{
		\mathcal A
	}
\newcommand{\mP} 
	{
		\mathcal P
	}
\newcommand{\N} 
{
	\mathcal N
}

\newcommand{\V} 
{
	\mathcal V
}
\newcommand{\G} 
{
	\mathcal G
}

\title{Identifying symmetries and predicting cluster synchronization in complex networks}

\author{Pitambar Khanra$^{1}$}
\thanks{These Authors equally contributed to the Manuscript}

\author{Subrata Ghosh$^{2}$}
\thanks{These Authors equally contributed to the Manuscript}

\author{Karin Alfaro-Bittner$^{3}$}

\author{Prosenjit Kundu$^{4}$}

\author{Stefano Boccaletti$^{5,6,7}$}

\author{Chittaranjan Hens$^{2}$}

\author{Pinaki Pal$^{1}$}

\affiliation{\noindent \textit{$^{1}$Department of Mathematics, National Institute of Technology, Durgapur 713209, India}}

\affiliation{\noindent \textit{$^{2}$Physics and Applied Mathematics Unit, Indian Statistical Institute, 203 B. T. Road, Kolkata 700108, India}}

\affiliation{\noindent \textit{$^{3}$Departamento de F\'isica, Universidad T\'ecnica Federico Santa Mar\'ia, Av. Espa{\~n}a 1680, Casilla 110V,  Valpara\'iso, Chile}}

\affiliation{\noindent \textit{$^{4}$Department of Mathematics, University at Buffalo, State University of New York, Buffalo, USA}}

\affiliation{\noindent \textit{$^{5}$Moscow Institute of Physics and Technology, National Research University, Dolgoprudny, Moscow, Russia}}

\affiliation{\noindent \textit{$^{6}$Universidad Rey Juan Carlos, Calle Tulip\'an s/n, 28933 M\'ostoles, Madrid, Spain}}

\affiliation{\noindent \textit{$^{7}$CNR - Institute of Complex Systems, Sesto Fiorentino, Italy}}
\date{\today}

\begin{abstract}
Symmetries in a network connectivity regulate how the graph's functioning organizes into clustered states.
Classical methods for tracing the symmetry group of a  network require very high computational costs, and therefore they are of hard, or even impossible, execution for large sized graphs.
We here unveil that there is a direct connection between the elements of the eigen-vector centrality and the clusters of a network. This gives a fresh framework for cluster analysis in undirected and connected graphs, whose computational cost is linear in $N$. We show that the
cluster identification is in perfect agreement with symmetry based analyses, and it allows predicting the sequence of synchronized clusters which form before the eventual occurrence of global synchronization.
\end{abstract}

 \maketitle

Synchronization in dynamical networks is one of the most common collective properties emerging in real and man made systems, from power grids to neuronal firing  ~\cite{Strogatz_synchronization_book, BoccalettiPhysRep2006, Dorfler_SIAM2012, Motter_NatPhys2013,ashwin2016mathematical,Boccalettibook}. In particular, complete, or global, synchronization (GS) is the state where all the identical units of a system evolve in unison. Yet, in some instances GS is either undesirable (it actually corresponds to pathological states of the brain) or is not attained. Rather, the network organizes into structure dependent synchronization states, such as cluster (CS) or remote synchronization \cite{dahms2012cluster,Skardal_PRE2011,Nicosia_PRL2013,sorrentino2007network,williams2013experimental,Pecora_NatCom2014,Sorrentino_SciAdv2016,lodi2020analyzing,bergner2012remote,sorrentino2016approximate,gambuzza2019criterion,Siddique_PRE2018, Sorrentino_SciAdv2016, Wang_Chaos2019,karakaya2019fading,zhang2017incoherence}. In CS, a set of units form a synchronized cluster \cite{Nicosia_PRL2013,cho2017stable,Pecora_NatCom2014,DellaRossa_NatCom2020,Sorrentino_IEEE2020} with the rest of the network evolving asynchronously.  Swarms of animals, or synchronous states (within sub units) in power grids, or brain dynamics are relevant examples of such CS \cite{Sorrentino_SciAdv2016}. In graph theoretic perspective, these clusters are the {\it orbits} of the graph and are the ingredients of the associated symmetry groups. The stability of such synchronous clusters depends on the analysis of the symmetry group elements \cite{Pecora_NatCom2014}.

Identifying the symmetries of a network is, therefore, of paramount importance, as it allows predicting the details of how the system's functioning organizes into synchronized clusters. However, tracing  the entire symmetry group of a network is not an easy task. It involves determining all symmetry group elements by a brute-force checking of permutations, a procedure which requires a number of operations which is non polynomial with the number $N$ of nodes, and therefore of hard (or even impossible) execution for large sized graphs.

Against this backdrop, in this Letter we introduce a general method through which one can identify the {\it orbits or clusters} of a network without the need of exploring its symmetry group. In particular, we analytically prove that there is a direct connection between the elements of the eigen-vector centrality (EVC) and the clusters of the network. The EVC is the eigen-vector of the network's adjacency matrix corresponding to the largest eigen-value \cite{pradhan2020principal,Newman_network_book}, and therefore the cost of its evaluation {\it scales linearly} with $N$.
We then propose a novel framework to unveil clusters in a generic undirected, connected, and positive semidefinite graph, and show that the
clusters identified are in perfect agreement with the ones determined from the symmetry based analysis \cite{DellaRossa_NatCom2020}.
Moreover, we show that our method discloses information also on the system's functioning, in that it allows predicting the sequence of clusters
which synchronize before the eventual occurrence of GS, {\it  independently on the specific dynamical units which are forming the network}.
We indeed investigated CS in networks of coupled finite as well as infinite dimensional identical chaotic units~\cite{Wang_Chaos2019}, and observed in both cases that the  emerging clusters are those identified by our method.
Thus, our approach opens a new window for  cluster analysis in networks, by drastically simplifying the procedure and reducing the associated computational cost, as compared to the existing methods based on symmetry groups.

\begin{figure}
\includegraphics[height=!,width=0.49\textwidth]{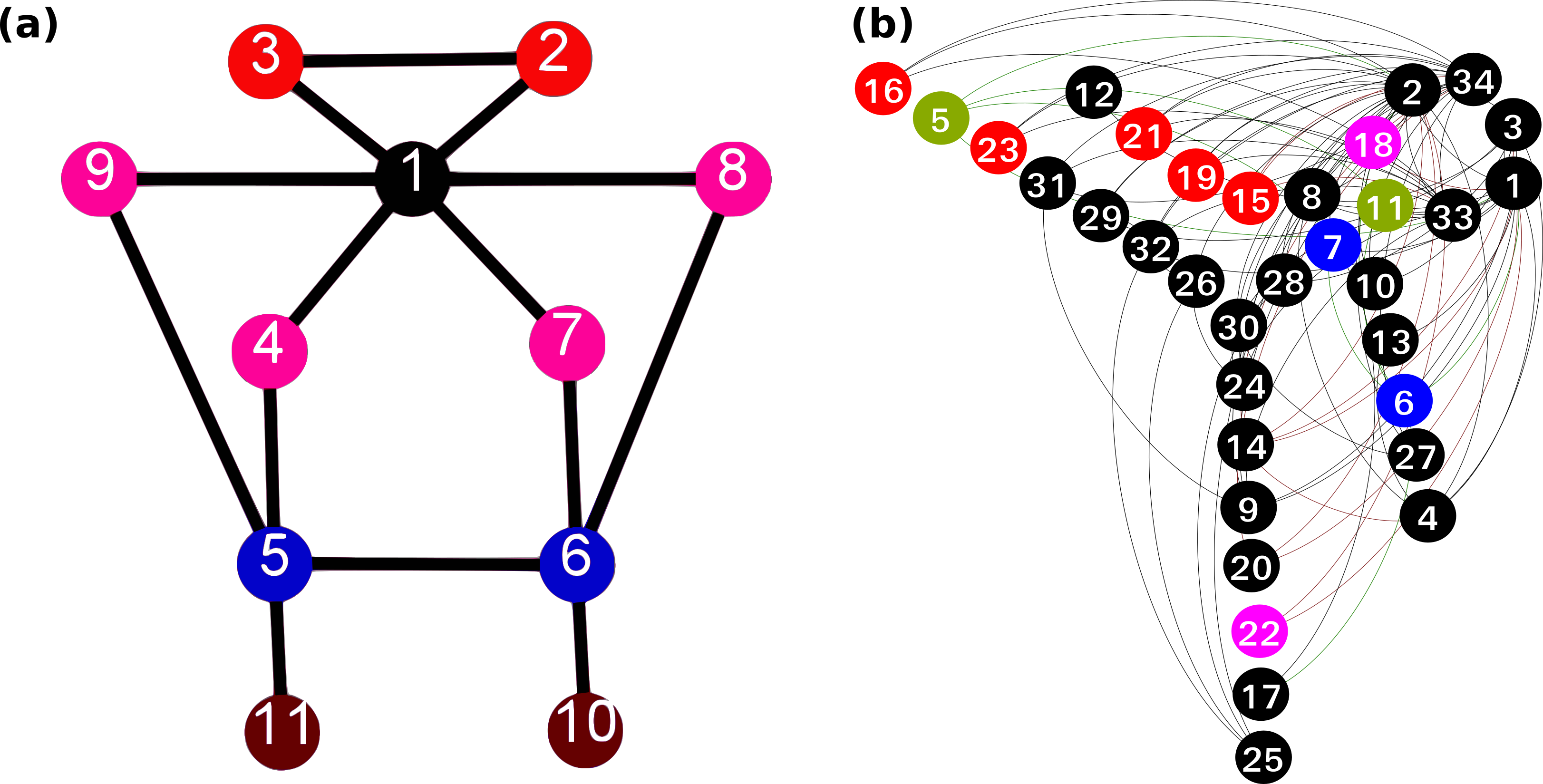}
\caption{{\bf Symmetries and clusters in the considered networks.} The graph in panel (a) is a synthetic network consisting of 11 nodes, and constructed in a way to display 4 sets of nontrivial clusters denoted by red, magenta, blue and brown color. The graph of panel (b) is, instead the well known Zachary karate club network \cite{zachary}, and consists of 34 nodes. Symmetry analysis identifies again 4 clusters, denoted by magenta, red, blue, and green color. Nodes colored in black do not participate in any cluster.}
\label{Figure1}
\end{figure}

Let us start by considering an undirected network $\mathcal G$ of size $\N$, and its adjacency matrix $\mathcal A = (A_{ij})_{\N\times \N}$.
Mathematically,  $\mathcal G$ is described as a connected graph $\G({\V},\mathcal{E})$, where ${\V}$ and $\mathcal{E}$ denote, respectively, the set of its vertices and edges, with $|{\V}|=\N$. Note that the edge set $\mathcal{E}$ consists of all ordered pairs $(i,j) \in {\V}\times {\V}$ such that the $i^{th}$ and $j^{th}$ nodes (vertices) are connected i.e. $\mathcal{E}\subseteq {\V} \times {\V}$.

Now, the graph $\mathcal G$ has a symmetry if and only if a bijective mapping $\Pi: {\V} \rightarrow {\V}$ exists that preserves the adjacency relation of $\mathcal G$, i.e. $\Pi$ is an automorphism for $\mathcal G$. In other words, a permutation matrix $\mathcal P=\mathcal P(\Pi)$ exists such that $\mathcal P \mathcal A (\mathcal P)^{-1} = \mathcal A$. The collection of all such $\mathcal P$ forms the symmetry group $G$ of the graph $\mathcal G$ with respect to matrix multiplication operation. The action of the group $G$ on the set of nodes ${\V}$ divides it  into disjoint invariant subsets which are called the {\it clusters or orbits} of the network.
Symmetries are directly connected to CS ~\cite{Nicosia_PRL2013}: the nodes within an orbit will synchronize also in absence of GS for suitable values of the system's parameters.

For instance, the graph in Fig.\ \ref{Figure1}(a) is a synthetic network of $11$ nodes. There exists $15$ non-trivial bijective mappings [see Sec.\ II of our supplementary material (SM)] which preserves the adjacency relation, and which (along with the identity mapping) forms a group under matrix multiplication.
One can then identify $4$ non-trivial orbits in this network: $(2,3: $ red nodes), $(5,6$ : blue nodes), $(9,8,4,7 :$ magenta nodes),   and $(10,11: $ brown nodes). Notice that the permutation of such nodes within an orbit preserves the connectivity pattern even when they are not neighbours to each other  (such as nodes 9 and 8).
The second network (Fig.\ \ref{Figure1}(b), $\N=34$) is the celebrated Zachary karate club network \cite{zachary}, and its group order (formed by $480$ elements) is increased by  a factor $\sim30$ as compared to the first case. Four non-trivial orbits/clusters can be  identified  from the group (see the SM for details on the generators and symmetry group of both networks). Depending on the structure of the graph, the number of  symmetries can be of the order of a million or a billion times the network size (see Table 1 in Ref. \cite{DellaRossa_NatCom2020}), thus making extremely complicated (if not impossible) the identification by classical methods of symmetry groups and orbits of large size networks.

The key question is then: can one extract the non-trivial clusters of an undirected graph without having to pass necessarily from the identification of the network's symmetry groups? Our answer is affirmative, and we will show that an enormously easier and less demanding operation (the inspection of the elements of the graph's EVC) is in fact sufficient for identifying the orbits and their member nodes. In the following, a proof is given which shows the direct relation between EVC and network clusters. \\

\begin{figure*}
\includegraphics[height=!,width=0.90\textwidth]{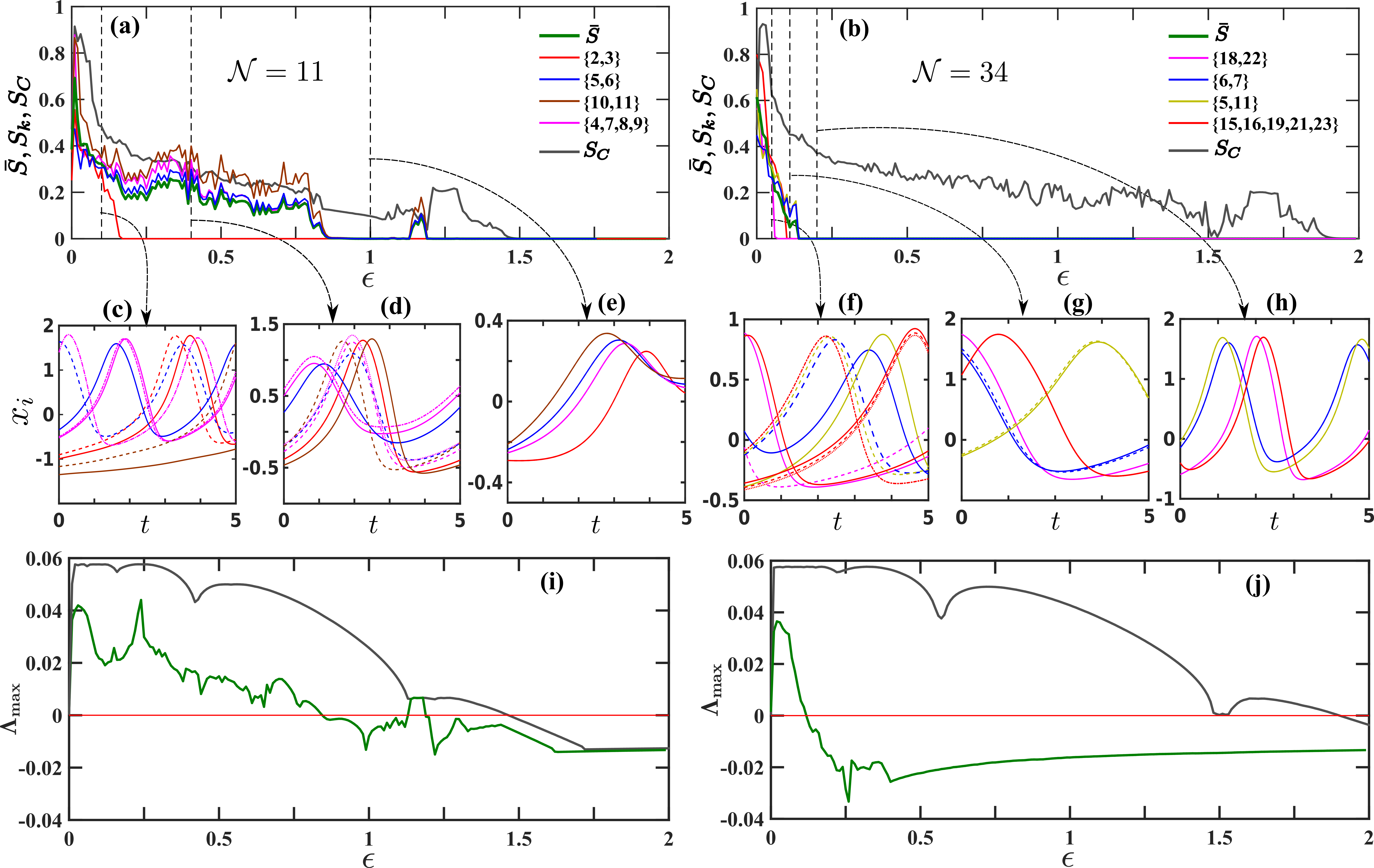}
\caption{{\bf Synchronization errors and transverse Lyapunov exponents.} (a,b): $\bar{S}, S_k, S_C$ (see text for definition) vs. $\epsilon$ for the two networks of Fig.\ (\ref{Figure1}). Color codes (reported in the legend of each panel) reflect the colors assigned to each cluster in Fig. (\ref{Figure1}).  We have  checked  three coupling regimes  (dashed vertical lines) for both networks. Panels (c-e) [(f-h)] report the time evolution of the nodes participating in each clusters of the first (the second) network, for the three values of $\epsilon$ marked with vertical dashed lines in panel (a) [(b)].  The full discussion on the different dynamical regimes observed is reported in the text. (i,j) The maximum Lyapunov exponents $\Lambda_{\rm max}$ transverse to the CCS (green) and the GS (black) manifolds vs. $\epsilon$.}
\label{Figure2}
\end{figure*}

Let $\mathbf{X}$ be the EVC, i.e. the eigen-vector of the  adjacency matrix $\mathcal A$ corresponding its largest eigen-value $\lambda_m$. As $\mathcal A$ is symmetric,  all its eigen-values are real, and the Perron-Frobenius theorem~\cite{Perron_1907theorie, frobenius1912matrizen} warrants that $\lambda_m$ is not degenerate, and that all the components of $\mathbf{X}$ are strictly positive. Then, one has $\mathcal A\mathbf{X}= \lambda_m \mathbf{X}$, and multiplying from left both sides by an element $\mathcal P$ of a symmetry group of $\mathcal{G}$ one gets $\mathcal{PA}\mathbf{X}=\lambda_m \mathcal P \mathbf{X}$, which implies $\mathcal{AP}\mathbf{X}=\lambda_m \mathcal P \mathbf{X}$ (because $\mathcal{AP = PA}$).

Assuming $\mathcal P \mathbf{X}=\mathbf{Y}$, one gets $\Rightarrow$ $\mathcal A \mathbf{Y}=\lambda_m \mathbf{Y}$.
Hence, along with $\mathbf{X}$, also $\mathbf{Y}$ is an eigenvector of $\mathcal A$ corresponding to the same eigenvalue $\lambda_m$. Thus, the set $\{\mathbf{X}, \mathbf{Y}\}$ must be linearly dependent, which implies the existence of a real number $c$ such that $\mathbf{Y}=c \mathbf{X}$, or $\mathcal P\mathbf{X} = c\mathbf{X}$.
But now, since $\mathbf{Y}$ is nothing but a rearrangement of the elements of $\mathbf{X}$, the value of $c$ must be $1$. Therefore, one has
\begin{equation}
\mathcal P\mathbf{X} = \mathbf{X}.\label{eqn8}
\end{equation}
Note that the relation (\ref{eqn8}) is true for all matrices $\mathcal P$ belonging to the symmetry group of $\mathcal{G}$, and it says that the EVC remains invariant under the action of such $\mathcal P$.
Now since each cluster of $\mathcal{G}$ is mapped to itself by the action of all such $\mathcal P$, Eq. (\ref{eqn8}) can hold {\it if and only if the components of $\mathbf{X}$ corresponding to the nodes of a cluster are equal}.
Moreover, the components of $\mathbf{X}$ corresponding to nodes of different clusters must be different. And this is because if they were equal then, apart from the symmetry group elements of the graph $\mathcal{G}$, there would be some other permutation matrices $\mathcal P'$ for which $\mathcal P'\mathbf{X} = \mathbf{X}$, and the action of those matrices on $\mathcal A$ would not satisfy $\mathcal{P'A} = \mathcal{AP'}$, with the immediate consequence that $\mathcal P'\mathbf{X} (= \mathbf{X})$ would not be an eigen-vector of $\mathcal A$. 

The simple and direct inspection of the EVC components allows, therefore, to identify the clusters of a network.

\par In what follows, we show that the clusters identified as the collections of those nodes displaying the same EVC value are exactly the ones where CS takes place,  independently on the specific dynamical system implemented in the networks' node. To do so, we consider a network of identical units, such that the uncoupled nodal dynamics is captured by $\dot{{\mathcal X}}= \mathbf F({\mathcal X})$, where dot denotes temporal derivative, ${\mathcal X}$ is a $p-$dimensional state vector, and ${\mathbf F}: {\Bbb R}^p \to {\Bbb R}^p$ is the flow function. The evolution equation for the $i^{th}$ unit ($i=1,...,N$)  is therefore
\begin{eqnarray}
\dot{\mathcal X}_i &=&\mathbf F({\mathcal X}_i)+ \mathbf E \sum _{j=1}^{\N}\A_{ij} \mathbf Q (\mathcal  X_j,  \mathcal X_i),
\label{model_eqn}
\end{eqnarray}
where $\mathbf E= \rm{diag}(\epsilon_1,\epsilon_2,\dots,\epsilon_p)$ is a $p \times p$ diagonal coupling matrix, and $\mathbf Q : {\Bbb R}^{2p} \to {\Bbb R}^p$ is a coupling function, which is here taken to be diffusive, i.e., $\mathbf Q (\mathcal  X_j, \mathcal X_i)=(\mathcal X_j-\mathcal X_i)$. The connectivity among the units is mapped into the $\N\times \N $ adjacency matrix $\A$.
Under a suitable choice of $\mathbf F$ and $\mathbf E$, the system may exhibit CS, e.g. the nodes within each orbit will synchronize their evolution.

Let us first discuss the properties of CS in a network of neurons, the dynamic of whose action potential is given by the Hindmarsh-Rose (HR) model \cite{Hindmarsh_Nature1982,Hindmarsh_PRS1984,Mishra_PRE2018dragon,Gonzalez_IJBC2007}. In the HR model, $p=3, \mathbf E= \rm{diag}(\epsilon,0,0), {\mathcal X}=(x,y,z)$, so that the evolution equation are ${\dot x}_{i}= y_i + bx_{i}^2 -ex_{i}^3 - z_{i} + I + \epsilon \sum_{i=1}^{\N}  \mathcal{A}_{ij} (x_j-x_i);  \ {\dot y}_{i} = c - d{x}_{i}^2 -y_{i}; \
{\dot z}_{i} = rs(x_{i}-x_{R})-rz_{i}.$
Here, $x$ is the membrane potential in the axon,  $y$ accounts for the fast ionic movement through membrane, and in the variable $z$ \cite{Hindmarsh_PRS1984} a slow exchange of  $Ca^{++}$ ions occurs due to the controlling parameter $r$. By properly setting the system's parameters (the external currents $I$ and $r$), several complex spiking and bursting pattern arise. Here, we consider $e=c=1$, $b=2.6$, $d=5$, $s=4$, $r=0.01$, and $x_{R}=-1.6$. Furthermore, all units are kept in their oscillatory state ($I=4$), and an electrical diffusive  coupling is introduced in the  variable $x$.
To monitor the onset of CS,  we refer to the quantity $\bar S$ defined by $\bar S=\left\langle S_k \right \rangle$,
where $\langle . \rangle$ denotes the average over the total number of non-trivial clusters, and $S_k$ is the time averaged root mean square deviation of the $k$th cluster  defined as $S_k=\left \langle \left(\frac{1}{\N_k}\sum_{i\in v_k}(x_i-\bar{x}_k)^2\right)^{1/2} \right \rangle_t$ (with $\N_k$ and $v_k$ being, respectively, the number and the set of nodes involved in the cluster $k$, and $\bar{x}_k$ denoting the average of the membrane potential of all axons within that cluster).

\begin{figure}
	\includegraphics[height=!,width=0.5\textwidth]{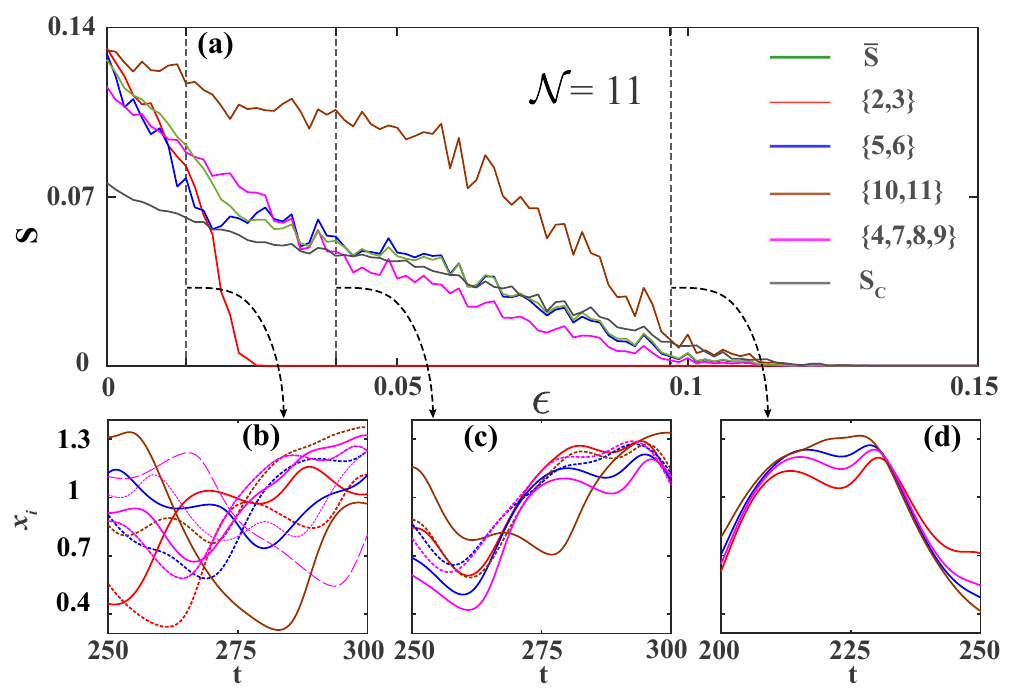}
	\caption{{\bf The Mackey-Glass system.} (a): $\bar{S}, S_k, S_C$ (see text for definition) vs. $\epsilon$ for the network of Fig.\ \ref{Figure1}(a). Color codes (reported in the legend) reflect the colors assigned to each cluster in Fig.~\ref{Figure1}. The black and green lines refer to $S_c$ and $\bar{S}$, respectively. (b-d): time series of each node forming clusters at three different values of $\epsilon$ (dashed vertical lines in panel (a)). Namely, $\epsilon = 0.0121$ (b), $\epsilon = 0.0394$ (c), and $\epsilon = 0.0970$ (d).}
	\label{Figure3}
\end{figure}

Extracting the identical elements of the EVC from the adjacency matrix of the graph sketched in Fig.\ \ref{Figure1}(a), it is fully confirmed that the network has four clusters. Namely, four nodes (marked by the magenta color in the Figure) participate in the largest orbital cluster, and the other three clusters are formed by two nodes each (marked by two blue nodes, two red nodes, and two brown nodes in the Figure). Fig.\ \ref{Figure2}(a) reports the overall synchronization scenario, as the coupling strength $\epsilon$ is increased. Four distinct regimes can be identified. For $\epsilon < 0.12$ the system stays in its fully incoherent state, which will be called from here on ``no synchronization" (NS). A typical behavior of the system is shown at $\epsilon=0.1$ [first dashed vertical line from left of Fig.\ \ref{Figure2}(a)] in Fig.\ \ref{Figure2}(c).
At $\epsilon\sim 0.12$, a first transition toward CS is seen, and a first cluster of nodes (nodes 2 and 3) synchronize. In Fig.\ \ref{Figure2}(d), we report the time signals at $\epsilon=0.4$ [second dashed vertical line from left in Fig.\ \ref{Figure2}(a)], and it is apparent that the thick red lines are indistinguishable, as the evolution of node 2 and 3 is synchronous. A second transition occurs at $\epsilon \sim 0.8$ towards Complete Cluster Synchronization (CCS).
There, the other three clusters synchronize almost together. CCS then corresponds to $\bar S =0$, where all clusters are separately synchronized, i.e. each of them evolves within a separated (non synchronized) dynamical state, as shown in Fig.\ \ref{Figure2}(e) for $\epsilon=1$ [third vertical line in Fig.\ \ref{Figure2}(a)]. Eventually,
the onset of GS, where  $\left(S_C=\left \langle \sqrt{\left\langle(x_i-\bar{x})^2\right\rangle_\N} \, \right \rangle_t \right) =0$, occurs at $\epsilon \sim 1.5$, and the entire network synchronize.
A similar scenario characterizes the dynamics of the Zachary karate club network, which has $\N=34$ nodes and $78$ links, and for which the EVC reveals again the presence of four nontrivial clusters [colored with magenta, blue, green, and red in Fig. \ref{Figure1}(b)]. Precisely, in the NS regime ($\epsilon=0.05$, Fig.\ \ref{Figure2}(b), left dashed vertical line)  all nodes are unsynchronized (see Fig.\ \ref{Figure2}(f) where all lines do not overlap). Increasing $\epsilon$ leads to CS. At the value of $\epsilon$ marked by the second dashed vertical line of Fig.\ \ref{Figure2}(b), two clusters (magenta and red nodes)  fire in unison, although with separated dynamics (Fig.\ \ref{Figure2}(g)). Eventually, all the orbits are synchronized ($\bar S=0$), i.e., CCS appears at $\epsilon_c=0.14$. The time signals at  $\epsilon=0.2$ are shown in Fig.\ \ref{Figure2}(h). GS ($S_c=0$) appears here at a much higher coupling strength, as it can be seen in Fig.\ \ref{Figure2}(b).

While assessment of stability of GS can be obtained from the Master Stability Function approach \cite{Pecora_PRL1998,Pecora_PRL1990,BoccalettiPhysRep2006}, and of that of CS by a parallel analysis of synchronization in each cluster \cite{Pecora_NatCom2014}, the conditions for stability of CCS are reported in details in our SM.
The maximum Lyapunov exponents $\Lambda_{\rm max}$ transverse to the CCS [green curves, from Eq.~(3) of the SM] and to the GS [black curves, from Eq.~(9) of the SM] manifolds are reported vs. $\epsilon$ in panels (i,j) of Fig. \ref{Figure2} for the two networks of Fig. \ref{Figure1}.
When compared with panels (a) and (b) of Fig. \ref{Figure2}, there is a perfect correspondence between the vanishing of these two exponents and the transition points to CCS and GS, respectively. Notice that the stability of GS occurs always later than that of CCS.

Remarkably, all qualitative features of the synchronization scenario are independent on the specific dynamical system implemented in each of the network nodes. For instance, we considered a network of infinite-dimensional systems, the delayed Mackey-Glass systems \cite{mackeyglass}, which obeys $\dot{x}_i(t) = -0.1 x_i(t) + 0.2\frac{x_i(t-\tau)}{1+x_i(t-\tau)^{10}}+ \epsilon \sum_{j=1}^{N} \A_{ij}\left[ x_j(t) - x_i(t)\right]$, where $\tau = 25$ is a delay time. Depending on the value of $\tau$ ($\tau > 16.8$), each node evolves in a hyper-chaotic attractor, and the dimensionality of the phase space scales linearly with $\tau$ \cite{jdfarmer}.
Results are reported in Figure \ref{Figure3} for the network of Fig.~\ref{Figure1}(a). In Fig.~\ref{Figure3}(a) it is clearly seen that the order at which each cluster synchronizes is exactly the same as the one shown in Fig.~\ref{Figure2}(a), i.e., the cluster marked in red {2,3} synchronizes first and the other three synchronize together at a higher value of the coupling strength. The overall scenario of synchronization is qualitatively the same as the one obtained for HR systems: a first transition from NS [Fig.~\ref{Figure3}(b)] to CS [Fig.~\ref{Figure3}(c)] is observed, but here the transition to CCS coincide with that to GS, and therefore the point at which all clusters synchronize is almost indistinguishable from that where network synchronization takes place [Fig.~\ref{Figure3}(d)].

In summary, we introduced a new framework for cluster analysis in undirected  positive semi-definite networks, based on the proof of a direct relation between the components of the EVC and the membership of each clusters or orbits of the network. This substantially simplifies cluster analysis of networks of any size, and drastically reduces its computational cost.  We further demonstrated that our analysis actually predicts the scenario of cluster synchronization supported by the network's connectivity, independently on the specific dynamical system implemented in the network nodes. Thus, it is easy to expect that our approach will empower the study of symmetry related problems in all cases and circumstances in which it was prevented so far by the computational costs associated to the classical methods, such as, for instance, in epileptic seizures, power-grid failures, or emerging clustered behavior in social networks and/or brain dynamics.

C.H. is supported by INSPIRE-Faculty grant (Code: IFA17-PH193).

\bibliographystyle{apsrev4-1} 

%

\end{document}


\newcommand{\I} 
	{
		\mathcal I
	}
\newcommand{\A} 
	{
		\mathcal A
	}
\newcommand{\mP} 
	{
		\mathcal P
	}
\newcommand{\N} 
{
	\mathcal N
}

\newcommand{\V} 
{
	\mathcal V
}
\newcommand{\G} 
{
	\mathcal G
}
	
\title{Supplementary Material: ``Identifying symmetries and predicting cluster synchronization in complex networks''}
\author{Pitambar Khanra}
\thanks{These Authors equally contributed to the Manuscript}

\author{Subrata Ghosh}
\thanks{These Authors equally contributed to the Manuscript}

\author{Karin Alfaro-Bittner}

\author{Prosenjit Kundu}

\author{Stefano Boccaletti}

\author{Chittaranjan Hens}

\author{Pinaki Pal}



\maketitle


In the following a series of details are reported about the networks considered in the main text, as well about the stability properties of the Global and Complete Cluster Synchronization manifolds, which led to the calculation of the two maximum Lyapunov exponents which are reported in Figs. 2(i,j) of the main text.

\section{Adjacency matrices corresponding to the Fig.\ 1 of the main text}
The adjacency matrices of the networks shown in the Fig.\ 1 of the main text are reported here below.  The corresponding eigen-vector centralities (EVC) are also shown in Equations (\ref{Bat_Net}) - (\ref{Karate_Club}). The EVC components associated with each nontrivial cluster are written with colored characters, while the black colored EVC components correspond to trivial or single node clusters. The color coding used in the EVC components are consistent with the color coding used in Fig.\ 1 of the main text.

\begin{itemize}

\item  Adjacency matrix of the first network ($\mathcal {N} = 11 $) and corresponding EVC:
\begin{eqnarray}\label{Bat_Net}
	\mathcal A=
	\begin{pmatrix}
	0 & 1 & 1 & 1 & 0 & 0 & 1 & 1 & 1 & 0 & 0\\
	1 & 0 & 1 & 0 & 0 & 0 & 0 & 0 & 0 & 0 & 0\\
	1 & 1 & 0 & 0 & 0 & 0 & 0 & 0 & 0 & 0 & 0\\
	1 & 0 & 0 & 0 & 1 & 0 & 0 & 0 & 0 & 0 & 0\\
	0 & 0 & 0 & 1 & 0 & 1 & 0 & 0 & 1 & 0 & 1\\
	0 & 0 & 0 & 0 & 1 & 0 & 1 & 1 & 0 & 1 & 0\\
	1 & 0 & 0 & 0 & 0 & 1 & 0 & 0 & 0 & 0 & 0\\
	1 & 0 & 0 & 0 & 0 & 1 & 0 & 0 & 0 & 0 & 0\\
	1 & 0 & 0 & 0 & 1 & 0 & 0 & 0 & 0 & 0 & 0\\
	0 & 0 & 0 & 0 & 0 & 1 & 0 & 0 & 0 & 0 & 0\\
	0 & 0 & 0 & 0 & 1 & 0 & 0 & 0 & 0 & 0 & 0
	\end{pmatrix},~~
	\textbf{EVC} =
	\begin{pmatrix}
	\color{black}0.5447676574199845634\\
	\color{red}0.2640912071035914077\\
	\color{red}0.2640912071035914077\\
	\color{magenta}{0.2850806807525254061}\\
	\color{blue}0.3283790981592057956\\
	\color{blue}0.3283790981592057956\\
	\color{magenta}{0.2850806807525254061}\\
	\color{magenta}0.2850806807525254061\\
	\color{magenta}0.2850806807525254061\\
	\color{brown}0.1072152641214826480\\
	\color{brown}0.1072152641214826480
	\end{pmatrix},		
\end{eqnarray}

\item  Adjacency matrix of the second network ($\N=34$) and corresponding EVC:
\footnotesize{
\begin{eqnarray}\label{Karate_Club}
	\mathcal A=
	\begin{pmatrix}
		0& 1& 1& 1& 1& 1& 1& 1& 1& 0& 1& 1& 1& 1& 0& 0& 0& 1& 0& 1& 0& 1& 0& 0& 0& 0& 0& 0& 0& 0& 0& 1& 0& 0\\
		1& 0& 1& 1& 0& 0& 0& 1& 0& 0& 0& 0& 0& 1& 0& 0& 0& 1& 0& 1& 0& 1& 0& 0& 0& 0& 0& 0& 0& 0& 1& 0& 0& 0\\
		1& 1& 0& 1& 0& 0& 0& 1& 1& 1& 0& 0& 0& 1& 0& 0& 0& 0& 0& 0& 0& 0& 0& 0& 0& 0& 0& 1& 1& 0& 0& 0& 1& 0\\
		1& 1& 1& 0& 0& 0& 0& 1& 0& 0& 0& 0& 1& 1& 0& 0& 0& 0& 0& 0& 0& 0& 0& 0& 0& 0& 0& 0& 0& 0& 0& 0& 0& 0\\
		1& 0& 0& 0& 0& 0& 1& 0& 0& 0& 1& 0& 0& 0& 0& 0& 0& 0& 0& 0& 0& 0& 0& 0& 0& 0& 0& 0& 0& 0& 0& 0& 0& 0\\
		1& 0& 0& 0& 0& 0& 1& 0& 0& 0& 1& 0& 0& 0& 0& 0& 1& 0& 0& 0& 0& 0& 0& 0& 0& 0& 0& 0& 0& 0& 0& 0& 0& 0\\
		1& 0& 0& 0& 1& 1& 0& 0& 0& 0& 0& 0& 0& 0& 0& 0& 1& 0& 0& 0& 0& 0& 0& 0& 0& 0& 0& 0& 0& 0& 0& 0& 0& 0\\
		1& 1& 1& 1& 0& 0& 0& 0& 0& 0& 0& 0& 0& 0& 0& 0& 0& 0& 0& 0& 0& 0& 0& 0& 0& 0& 0& 0& 0& 0& 0& 0& 0& 0\\
		1& 0& 1& 0& 0& 0& 0& 0& 0& 0& 0& 0& 0& 0& 0& 0& 0& 0& 0& 0& 0& 0& 0& 0& 0& 0& 0& 0& 0& 0& 1& 0& 1& 1\\
		0& 0& 1& 0& 0& 0& 0& 0& 0& 0& 0& 0& 0& 0& 0& 0& 0& 0& 0& 0& 0& 0& 0& 0& 0& 0& 0& 0& 0& 0& 0& 0& 0& 1\\
		1& 0& 0& 0& 1& 1& 0& 0& 0& 0& 0& 0& 0& 0& 0& 0& 0& 0& 0& 0& 0& 0& 0& 0& 0& 0& 0& 0& 0& 0& 0& 0& 0& 0\\
		1& 0& 0& 0& 0& 0& 0& 0& 0& 0& 0& 0& 0& 0& 0& 0& 0& 0& 0& 0& 0& 0& 0& 0& 0& 0& 0& 0& 0& 0& 0& 0& 0& 0\\
		1& 0& 0& 1& 0& 0& 0& 0& 0& 0& 0& 0& 0& 0& 0& 0& 0& 0& 0& 0& 0& 0& 0& 0& 0& 0& 0& 0& 0& 0& 0& 0& 0& 0\\
		1& 1& 1& 1& 0& 0& 0& 0& 0& 0& 0& 0& 0& 0& 0& 0& 0& 0& 0& 0& 0& 0& 0& 0& 0& 0& 0& 0& 0& 0& 0& 0& 0& 1\\
		0& 0& 0& 0& 0& 0& 0& 0& 0& 0& 0& 0& 0& 0& 0& 0& 0& 0& 0& 0& 0& 0& 0& 0& 0& 0& 0& 0& 0& 0& 0& 0& 1& 1\\
		0& 0& 0& 0& 0& 0& 0& 0& 0& 0& 0& 0& 0& 0& 0& 0& 0& 0& 0& 0& 0& 0& 0& 0& 0& 0& 0& 0& 0& 0& 0& 0& 1& 1\\
		0& 0& 0& 0& 0& 1& 1& 0& 0& 0& 0& 0& 0& 0& 0& 0& 0& 0& 0& 0& 0& 0& 0& 0& 0& 0& 0& 0& 0& 0& 0& 0& 0& 0\\
		1& 1& 0& 0& 0& 0& 0& 0& 0& 0& 0& 0& 0& 0& 0& 0& 0& 0& 0& 0& 0& 0& 0& 0& 0& 0& 0& 0& 0& 0& 0& 0& 0& 0\\
		0& 0& 0& 0& 0& 0& 0& 0& 0& 0& 0& 0& 0& 0& 0& 0& 0& 0& 0& 0& 0& 0& 0& 0& 0& 0& 0& 0& 0& 0& 0& 0& 1& 1\\
		1& 1& 0& 0& 0& 0& 0& 0& 0& 0& 0& 0& 0& 0& 0& 0& 0& 0& 0& 0& 0& 0& 0& 0& 0& 0& 0& 0& 0& 0& 0& 0& 0& 1\\
		0& 0& 0& 0& 0& 0& 0& 0& 0& 0& 0& 0& 0& 0& 0& 0& 0& 0& 0& 0& 0& 0& 0& 0& 0& 0& 0& 0& 0& 0& 0& 0& 1& 1\\
		1& 1& 0& 0& 0& 0& 0& 0& 0& 0& 0& 0& 0& 0& 0& 0& 0& 0& 0& 0& 0& 0& 0& 0& 0& 0& 0& 0& 0& 0& 0& 0& 0& 0\\
		0& 0& 0& 0& 0& 0& 0& 0& 0& 0& 0& 0& 0& 0& 0& 0& 0& 0& 0& 0& 0& 0& 0& 0& 0& 0& 0& 0& 0& 0& 0& 0& 1& 1\\
		0& 0& 0& 0& 0& 0& 0& 0& 0& 0& 0& 0& 0& 0& 0& 0& 0& 0& 0& 0& 0& 0& 0& 0& 0& 1& 0& 1& 0& 1& 0& 0& 1& 1\\
		0& 0& 0& 0& 0& 0& 0& 0& 0& 0& 0& 0& 0& 0& 0& 0& 0& 0& 0& 0& 0& 0& 0& 0& 0& 1& 0& 1& 0& 0& 0& 1& 0& 0\\
		0& 0& 0& 0& 0& 0& 0& 0& 0& 0& 0& 0& 0& 0& 0& 0& 0& 0& 0& 0& 0& 0& 0& 1& 1& 0& 0& 0& 0& 0& 0& 1& 0& 0\\
		0& 0& 0& 0& 0& 0& 0& 0& 0& 0& 0& 0& 0& 0& 0& 0& 0& 0& 0& 0& 0& 0& 0& 0& 0& 0& 0& 0& 0& 1& 0& 0& 0& 1\\
		0& 0& 1& 0& 0& 0& 0& 0& 0& 0& 0& 0& 0& 0& 0& 0& 0& 0& 0& 0& 0& 0& 0& 1& 1& 0& 0& 0& 0& 0& 0& 0& 0& 1\\
		0& 0& 1& 0& 0& 0& 0& 0& 0& 0& 0& 0& 0& 0& 0& 0& 0& 0& 0& 0& 0& 0& 0& 0& 0& 0& 0& 0& 0& 0& 0& 1& 0& 1\\
		0& 0& 0& 0& 0& 0& 0& 0& 0& 0& 0& 0& 0& 0& 0& 0& 0& 0& 0& 0& 0& 0& 0& 1& 0& 0& 1& 0& 0& 0& 0& 0& 1& 1\\
		0& 1& 0& 0& 0& 0& 0& 0& 1& 0& 0& 0& 0& 0& 0& 0& 0& 0& 0& 0& 0& 0& 0& 0& 0& 0& 0& 0& 0& 0& 0& 0& 1& 1\\
		1& 0& 0& 0& 0& 0& 0& 0& 0& 0& 0& 0& 0& 0& 0& 0& 0& 0& 0& 0& 0& 0& 0& 0& 1& 1& 0& 0& 1& 0& 0& 0& 1& 1\\
		0& 0& 1& 0& 0& 0& 0& 0& 1& 0& 0& 0& 0& 0& 1& 1& 0& 0& 1& 0& 1& 0& 1& 1& 0& 0& 0& 0& 0& 1& 1& 1& 0& 1\\
		0& 0& 0& 0& 0& 0& 0& 0& 1& 1& 0& 0& 0& 1& 1& 1& 0& 0& 1& 1& 1& 0& 1& 1& 0& 0& 1& 1& 1& 1& 1& 1& 1& 0

	\end{pmatrix} \label{multi_Fig1},
	\textbf{EVC} =
	\begin{pmatrix}
		\color{black}0.355490721123704\\
		\color{black}0.265959604787813\\
		\color{black}0.317192417268809\\
		\color{black}0.211179693646466\\
		\color{olive}0.075968878668705\\
		\color{blue}0.079483113040960\\
		\color{blue}0.079483113040960\\
		\color{black}0.170959892270922\\
		\color{black}0.227404355331627\\
		\color{black}0.102674503621447\\
		\color{olive}0.075968878668705\\
		\color{black}0.052855816844280\\
		\color{black}0.084254726652952\\
		\color{black}0.226473112020524\\
		\color{red}0.101403649968358\\
		\color{red}0.101403649968358\\
		\color{black}0.023635566309504\\
		\color{magenta}0.092399698590404\\
		\color{red}0.101403649968358\\
		\color{black}0.147912918340006\\
		\color{red}0.101403649968358\\
		\color{magenta}0.092399698590404\\
		\color{red}0.101403649968358\\
		\color{black}0.150118911532114\\
		\color{black}0.057052325511109\\
		\color{black}0.059206342354626\\
		\color{black}0.075579615970571\\
		\color{black}0.13347738580775\\
		\color{black}0.13107796378977\\
		\color{black}0.134961122254987\\
		\color{black}0.174758637204661\\
		\color{black}0.191034394033312\\
		\color{black}0.308643748591797\\
		\color{black}0.373362538968214
	\end{pmatrix}, \nonumber \\		
\end{eqnarray}}

\end{itemize}

\section{Symmetry group corresponding to the Fig. 1(a) of the main text}
The graph in Fig. 1(a) of the main text has one trivial bijective mapping, i.e. the identity mapping,
and 15 non-trivial bijective mapping which preserves the adjacency relations. These mappings are:
\begin{eqnarray}
\Pi_1 : (1,2,3,4,5,6,7,8,9,10,11) \to (1,3,2,4,5,6,7,8,9,10,11) \nonumber\\
\Pi_2 : (1,2,3,4,5,6,7,8,9,10,11) \to (1,2,3,9,5,6,7,8,4,10,11) \nonumber\\
\Pi_3 : (1,2,3,4,5,6,7,8,9,10,11) \to (1,2,3,4,5,6,8,7,9,10,11) \nonumber\\
\Pi_4 : (1,2,3,4,5,6,7,8,9,10,11) \to (1,2,3,7,6,5,4,9,8,11,10) \nonumber\\
\Pi_5 : (1,2,3,4,5,6,7,8,9,10,11) \to (1,2,3,8,6,5,4,9,7,11,10) \nonumber\\
\Pi_6 : (1,2,3,4,5,6,7,8,9,10,11) \to (1,2,3,9,5,6,8,7,4,10,11) \nonumber\\
\Pi_7 : (1,2,3,4,5,6,7,8,9,10,11) \to (1,2,3,7,6,5,9,4,8,11,10) \nonumber\\
\Pi_8 : (1,2,3,4,5,6,7,8,9,10,11) \to (1,2,3,8,6,5,9,4,7,11,10) \nonumber\\
\Pi_9 : (1,2,3,4,5,6,7,8,9,10,11) \to (1,3,2,4,5,6,8,7,9,10,11) \nonumber\\
\Pi_{10} : (1,2,3,4,5,6,7,8,9,10,11) \to (1,3,2,7,6,5,4,9,8,11,10) \nonumber\\
\Pi_{11} : (1,2,3,4,5,6,7,8,9,10,11) \to (1,3,2,8,6,5,4,9,7,11,10) \nonumber\\
\Pi_{12} : (1,2,3,4,5,6,7,8,9,10,11) \to (1,3,2,9,5,6,7,8,4,10,11) \nonumber\\
\Pi_{13} : (1,2,3,4,5,6,7,8,9,10,11) \to (1,3,2,9,5,6,8,7,4,10,11) \nonumber\\
\Pi_{14} : (1,2,3,4,5,6,7,8,9,10,11) \to (1,3,2,7,6,5,9,4,8,11,10) \nonumber\\
\Pi_{15} : (1,2,3,4,5,6,7,8,9,10,11) \to (1,3,2,8,6,5,9,4,7,11,10) \nonumber
\end{eqnarray}

The network has four generators: \\
 $[(2,3),\\
(4,9),\\
(7,8),\\
(4,7)(5,6)(8,9)(10,11)] $. \\

In case of the graph of Fig. 1(b), the order of the symmetry group is 480, and the underlying generators are: \\
$[(21,23),\\
(19,21),\\
(18,22),\\
(16,19),\\
(15,16),\\
(5,11)(6,7)]$.

A close inspection on generators gives the nontrivial-clusters of each network.

\section{Stability of the Complete Cluster synchronization manifold}

We start from Eq. (2) of the main text, and notice that, in a purely diffusive formalism, such an Equation can be rewritten as
$$\dot{\mathcal X}_i =\mathbf F({\mathcal X}_i,t)- \mathbf E\sum _{j=1}^{\N} \mathcal L_{ij} \mathbf H (\mathcal  X_j),~~i=1,\dots,\mathcal{N},$$
where $\mathcal L_{ij}$ is the Laplacian matrix corresponding to the adjacency matrix $\mathcal A_{ij}$, and for our specific HR model one further has
 $\mathbf H(\mathcal X_i)=\mathcal X_i$.

Now, $\mathcal{G}$ can be partitioned into $\wp$ number of orbital clusters, namely $\mathcal{C}_1,~\mathcal{C}_2,\dots,\mathcal{C}_{\wp}$.
One further defines the $\wp\times\wp$ dimensional quotient matrix $\mathcal{Q}$ such that, for each pair of orbital clusters $(\mathcal{C}_u,\mathcal{C}_v)$.
One then has
$$\mathcal{Q}_{uv}= \sum_{j\in \mathcal{C}_v}\mathcal L_{ij}~;~~~~~\forall i\in\mathcal{C}_u,~~~u,v=1,2,\dots,\wp. $$

Let $\mathcal{S}_u$ be the synchronization manifold corresponding to the clusters $\mathcal{C}_u$,~$u=1,2,\dots,\wp$. Thus, one can write
$$\dot {\mathcal{S}}_u = \mathbf F({\mathcal{S}}_u) - \mathbf E \sum_{v=1}^{\wp}\mathcal{Q}_{uv} \mathbf{H}(\mathcal{S}_v).$$

CCS corresponds to the situation where $\mathcal X_i\equiv \mathcal{S}_u$, $\forall i \in \mathcal{C}_u$. One then considers small perturbations $\delta{\mathcal X_i}$ around the synchronized manifold, so that $\mathcal X_i=\mathcal{S}_u+\delta{\mathcal X_i}$, $\forall i \in \mathcal{C}_u$, and linear stability analysis gives a variational equation

\begin{eqnarray}
\dot \delta \mathcal X_i = \mathbf{JF} ({\mathcal{S}}_u) \delta \mathcal X_i - \mathbf E \sum_{j=1}^{N} \mathcal L_{ij} \mathbf{JH}(\mathcal{S}_v)\delta \mathcal X_j,
\end{eqnarray}
where $\mathbf{JF}$, and $\mathbf{JH}$ are the Jacobians of $\mathbf F$ and $\mathbf{H}$. Such a latter equation allows to calculate the maximal Lyapunov exponent $\Lambda_{\rm max}$ along the transverse direction to CCS: a negative value of it denotes stability of the CCS regime.

\section{Stability of the Global synchronization manifold}
Let $\mathcal S$ indicates the global synchronization (GS) manifold, which is a hyperplane in the phase space. The time evolution equation for the synchronization manifold can be written as,
\begin{eqnarray}\label{manifold}
\dot{\mathcal S}= \mathbf F(\mathcal S)-\mathbf E \sum _{j=1}^{\N} \mathcal L_{ij} \mathbf H (\mathcal  S)
\end{eqnarray}
$\sum_{j=1}^\mathcal N \mathcal L_{ij}=0,~ \forall i=1,\dots,\mathcal N$ property ensures the solution $\dot{\mathcal S}= \mathbf F(\mathcal S)$. Every node $\mathcal X_i$ of node $i$ is mapped into the GS manifold $\mathcal S(x_s,y_s,z_s)$ such that $\mathcal X_1=\mathcal X_2=\dots=\mathcal X_\mathcal N=\mathcal S$. Now considering small perturbations $\delta{\mathcal X_i}$ around the GS manifold, the phase becomes $\mathcal X_i=\mathcal{S}+\delta{\mathcal X_i}$, $\forall i$. Then the variational equation corresponding to Eq.\ (11) of the main text is,
\begin{eqnarray}\label{perturbed_eqn}
\dot \delta \mathcal X_i = \mathbf{JF} (\mathcal{S}) \delta \mathcal X_i - \mathbf E \sum_{j=1}^{N} \mathcal L_{ij} \mathbf{J}\mathbf{H}(\mathcal{S})\delta \mathcal X_j
\end{eqnarray}
where $\mathbf{JF}$, and $\mathbf{JH}$ denote the jacobian corresponding to $\mathbf F$, and the coupling function $\mathbf{H}$. Analysis of the stability of $\mathcal S$ is done by considering devisations along directions transverse to the $\mathcal S$ hyperplane. Therefore, one eventually has to solve an eigenvalues problem.\\
Let $\lambda_i$ and $\mathbf v_i$ be the eigenvalues and the eigenvectors corresponding to the laplacian matrix $\mathcal L$. So the perturbation vector $\delta \mathcal X_i$ can be expanded with the orthonormal basis formed by $\{\mathbf v_i\}$ as,
\begin{eqnarray}
\delta \mathcal X_i = \sum_{i=1}^\mathcal N \mathbf v_i \oplus \xi_i,
\end{eqnarray}
where $\xi_i$ is the new variable. Now applying $\mathbf v_i^T$ on both sides of the linearized equation Eq.\ (\ref{perturbed_eqn}) and substitute the expansion we can finally get the variational equation as,
\begin{eqnarray}\label{final_vari}
\dot \xi_i = [\mathbf{JF} (\mathcal{S}) - \lambda_i \mathbf E\mathbf{J}\mathbf{H}(\mathcal{S})]\xi_i.
\end{eqnarray}
The solution of the Eq.\ (\ref{final_vari}) will be a function of the eigenvalue $\lambda_i$, which is the required Master Stability  Function (MSF) of the GS manifold. For different network structures, the value of $\lambda_i$ is can be varied  but easily determined. That leads to the transversal Lyapunov Exponent as,
\begin{eqnarray}
\Lambda_i^j=\lim_{t \to \infty} \frac{\xi_i^j}{t},~j=1,2,3 ~\&~ i=1,\dots,\mathcal N.
\end{eqnarray}
The maximum of these lyapunov exponents is,
\begin{eqnarray}
\Lambda_{\rm {max}}=\rm{max}(\rm{max}(\Lambda_i^j,~j=1,2,3),~i=1,\dots,\mathcal N).
\end{eqnarray}
Negative (positive) values of that maximal tranversal Lyapunov exponent indicates stability (instability) of the GS manifold.